\documentclass{article}

\usepackage[final]{neurips_2019}
\setcitestyle{numbers}

\usepackage[utf8]{inputenc}
\usepackage[T1]{fontenc}
\usepackage{hyperref}
\usepackage{url}
\usepackage{booktabs}
\usepackage{amsfonts}
\usepackage{nicefrac}
\usepackage{microtype}
\usepackage{graphicx}
\usepackage{xcolor}
\usepackage{lipsum}
\usepackage{multirow}
\usepackage{graphicx}
\graphicspath{ {./images/} }
\usepackage{wrapfig}
 \usepackage{float}

\title{
  Power-of-Two (PoT) Weights in Large Language Models (LLMs) \\
  \vspace{1em}
}

\author{
  Mahmoud Elgenedy \\
  Department of Computer Science \\
  Stanford University \\
  \texttt{melgened@stanford.edu} \\
}

\begin{document}

\maketitle

\begin{abstract}
Complexity of Neural Networks is increasing rapidly due to the massive increase in model parameters. Specifically, in Large Language Models (LLMs), the number of model parameters has grown exponentially in the past few years, for example, from 1.5 billion parameters in GPT2 to 175 billion in GPT3. This raises a significant challenge for implementation, especially for Edge devices where memory and processing power are very limited. In this work, we investigate reducing LLM complexity with special type of quantization, power of two (PoT), for linear layers weights and transformer tables. PoT not only provides memory reduction but more importantly provides significant computational reduction through converting multiplication to bit shifting. We obtained preliminary results of PoT quantization on Nano-GPT implementation using Shakespeare dataset. We then extended results to 124-M GPT-2 model. The PoT quantization results are shown to be very promising with cross entropy loss degradation $\approx$[1.3-0.88] with number of bits range [4-6] to represent power levels.

\end{abstract}


\section{Introduction}
The unprecedented advancements in Large Language models (LLMs) are revolutionizing everything in our life. However, the cost of improvement is becoming increasingly expensive which limits this powerful tool in many applications, especially considering edge devices. Unlike cloud computing, memory and processing power for Edge devices are very limited, which necessitates developing efficient implementation to make such applications feasible. A small enhancement in implementation may result in millions of dollars saving and opens the door for many more applications\footnote{We are quoting the following statement of Andrew NG, commenting on the DeepSeek huge disruption to AI technology, “Scaling up isn't the only path to Al progress. Despite the massive focus on and hype around processing power, algorithmic innovations are rapidly pushing down training costs.”}.

One effective way of optimization is quantization, where numbers can be represented in lower precision to help reduce both memory and processing requirements. In this proposal, we are investigating compressing LLM model parameters through efficient quantization. Specifically, we are  leveraging the most advanced framework provided in PyTorch to investigate baseline normal quantization. Further, we explore more aggressive approach, power of two, which will not only help reducing memory requirements (as only exponents need to be stored) but more importantly provides incredible processing saving by converting multiplication into bit-shifting. For example, using 8-bits integer instead of floating 32 bits, provides 75\% saving of parameters storage. Imagine the savings when applying that to ~175 billion parameters in GPT3 or even higher (up to ~671 billion parameters as announced in the DeepSeek-R1 model). However, the drawback is the performance degradation which is quantified in this research.

\section{Related Work}
While quantization in neural networks is widely researched \citep{gholami2022survey,krishnamoorthi2018quantizing,wu2020integer}, only few studies investigated power-of-two quantization \citep{przewlocka2022power}. Specifically, restricting weights to PoT is investigated in a couple of studies for general Deep Neural Networks (DNNs) and with some focus on image applications using Convolutional Neural Networks (CNNs) \citep{mcdanel2019full}. However, for LLMs applications, there is no clear study for that approach.
Different studies on optimizing memory requirements for the weights in LLMs, including reducing bit-width to 4-bits as well as pruning algorithms \citep{Neves_2025}, suggesting feasibility of proposed PoT weights restriction.

\section{LLM Model and Dataset}
\subsection{LLM model (Nano-GPT)}
\begin{figure}[ht] 
    \caption{LLM Model illustration and parameters}
    \label{model}
    \centering
    \includegraphics[width=1\textwidth]{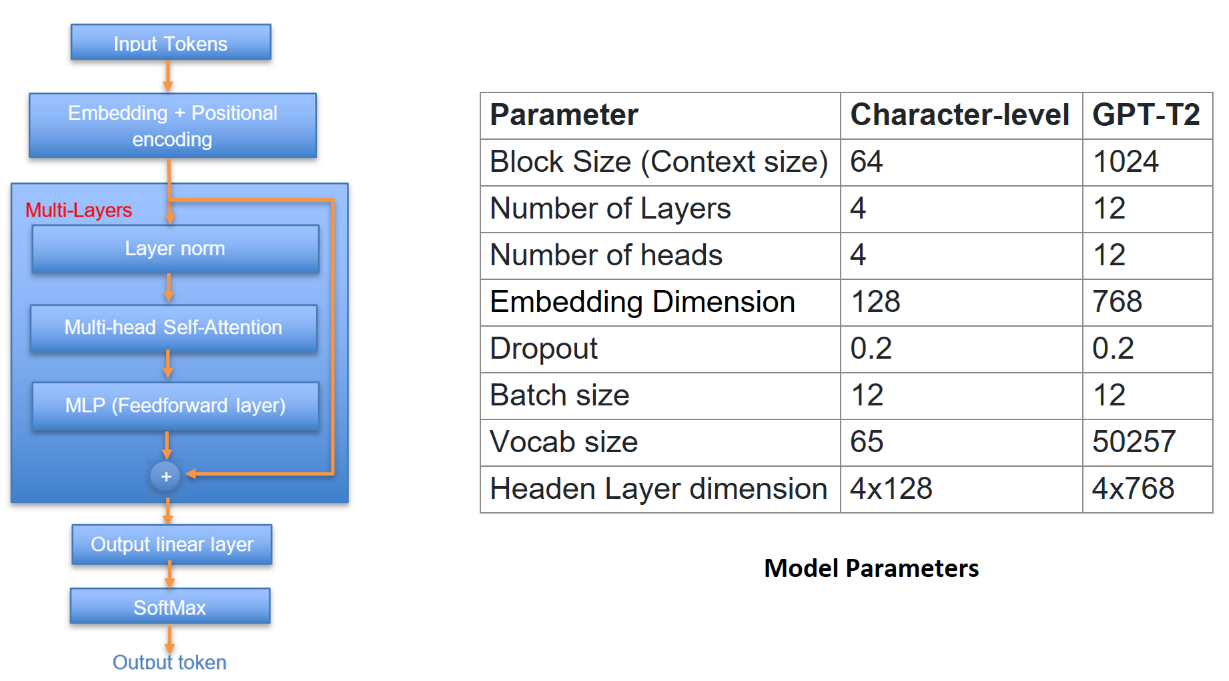}
\end{figure}

NanoGPT \cite{Karpathy1} is a simple repository for training and tuning Generative Pre-trained Transformer GPT. The design follows GPT-2 model and can reproduce GPT-2 on OpenWebText dataset \cite{openwebtext}. Moreover, it supports various settings including character-level LLM allowing fast trials and smaller datasets. Figure \ref{model} shows a simplified block diagram for the model with detailed parameters, while Table \ref{complexity-table}, details the complexity of simple character-level and GPT-2 models.

\subsection{Datasets}
We considered two main datasets, a small dataset for quick trials (Shakespeare dataset), and another large dataset for scalability testing (OpenWebText dataset).  A sample of text and equivalent encoded tokens for both datasets are shown in Appendix.
\begin{enumerate}
\item{\textbf{Shakespeare dataset}}\newline
A small dataset ($\approx 1$ MB) consists of texts from William Shakespeare's plays. Our preliminary trials considering a character-level small LLM model trained only on Shakespeare dataset. We partitioned the data as 80\%-10\%-10\% for training-val-testing. Character-level Tokenizer, from nano-GPT, is used to encode/decode dataset.
\item{\textbf{OpenWebText dataset}}\newline
An open-source recreation of OpenAI's WebText dataset, which mainly used to train GPT-2. It is relatively large ($\approx 54$ GB) dataset consists of news articles, blog posts and stories. We evaluated the performance of quantization on OpenWebText dataset using a pretrained GPT-2 model \citep{GPT2LMHeadModel}. The evaluation dataset is ~0.05\% of the total dataset. A Tokenizer from "tiktoken" \citep{tiktoken} is used.
\end{enumerate}

In pre-processing, algorithm loads the dataset, then encode it using the Tokenizer, and store in ".bin" files labeled as "train", "test", and "val". This saves the time during evaluation for different trials.



\begin{table}
\caption{LLM character-level (CL) and GPT-2 number of parameters and complexity (per layer)}
\label{complexity-table}

\begin{tabular}{ |p{3.0cm}||p{2.0cm}|p{2.5cm}|p{2.0cm}| p{2.8cm}| } 
 \hline
 Component& Parameters count (CL) &Complexity (CL)
&Parameters count (GPT-2)
&Complexity (GPT-2)
\\
 \hline
 Embedding Layer
   & 65x128    &LUT&   50257x768  & LUT\\
 Position Embeddings
&   64x128  & LUT   &1024x768 &LUT\\
 Multi-head Attention &4x128x128 & O(64x128x128) + O(64x64x128)
&  12x768x768 & O(1024x768x768) + O(1024x1024x768)\\
 MLP    &2x(128x4x128) & O(64x4x128x128)&  2x(768x4x768)& O(1024x4x768x768)\\
 Layer norm
&   2x128  & O(64x128)&2x768& O(1024x768)\\
 Output Linear Layer& 128x65  & O(128x65x64)   &768x50257& O(768x50257x1024)\\
 \hline
\end{tabular}
\end{table}

\section{Methods}
\subsection{PyTorch Quantization Framework}
Quantization is a technique to represent numbers in lower precision bit-widths (e.g., 8-bits) rather than full-precision floating point (32-bits). Model quantization can significantly reduce memory requirements. Moreover, quantization changes representation from floating point format (mantissa + exponent) to a simpler integer representation. Two main types of representation, symmetric versus asymmetric quantization. Symmetric quantization assumes zero offset of the data and only uses a scale to map real world numbers into integers. However, Asymmetric (affine) uses both a scale and zero-point (zp) to map the real world numbers into integers. Rounding operation is usually considered after scaling. In this proposal, we will assume another type, which is power-of-two quantization, that will limit integer numbers to be only power of two. Table \ref{quant-table} summarizes the three types.

PyTorch provides different quantization frameworks \citep{Quantization}, including Eager Mode, FX Graph Mode, and PyTorch 2 Export Mode. We choose to work with most recent PyTorch 2 Export Mode \citep{Pytorch} as it is more flexible and easily scalable. Both symmetric and Asymmetric are supported in PyTorch frameworks. However, PoT is not supported, and part of this work is to add it to PyTorch framework.
In addition, Pytorch also provides different criteria to quantize Deep learning models, Dynamic quantization, Static Post Training quantization, and Static Quantization Aware Training. In summary, Dynamic mode will quantize weights during the inference time (advantage is to adjust weights in real-time with added complexity). On the other hand, static quantization quantizes wights/activations offline but requires either calibration (Post Training Quantization PTQ) or training (Quantization Aware Training QAT). In this proposal, we assume only static quantization, PTQ.

\begin{table}
\caption{Types of Quantization}
\label{quant-table}

\begin{tabular}{ |p{3.0cm}||p{4.5cm}|p{5cm}| } 
 \hline
 Quantization type & Conversion equation (input x, and output y) & Use case\\
 \hline
 Symmetric
   & $y = round(x/scale)$    &Weights\\
 Asymmetric (Affine)
&   $y=round(x/scale) + zp$  & Activations\\
 Power of Two (PoT) &$ y = 2^{clip(round(\log_2(x/scale)))}$ & Suggested in this proposal for Weights and Transformer tables\\
 
 \hline
\end{tabular}
\end{table}

\subsection{Quantized Model}
\begin{figure}[ht] 
    \caption{PyTorch Quantization Framework}
    \label{quant}
    \centering
    \includegraphics[width=1\textwidth]{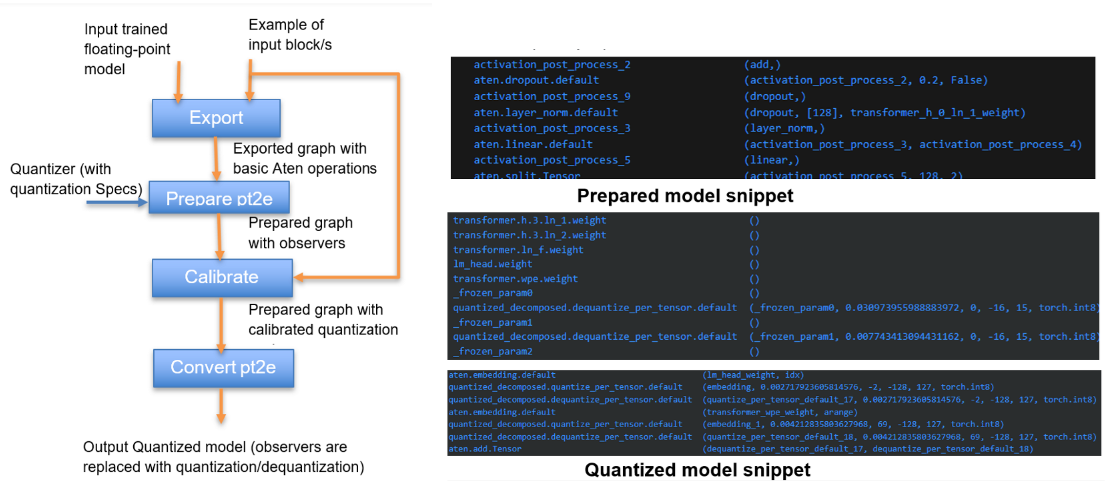}
\end{figure}

Figure \ref{quant} shows the PyTorch 2 Export quantization framework. The first step is to export the trained floating-point model into a graph basic tensor operations. Second, “Prepare Pt2e”, where observers are added before and after each operation (used to calibrate input and output range). The third step is to run the prepared model with realistic data so that observers can do calibration. Final step is to convert each observer into quantize/dequantize operation. Figure \ref{quant}, shows snippets of prepared and quantized model (Graph).
It is worth noting that generated quantized graph is not the final quantized model, hence it uses de-quantize operations. However, it should simulate the quantization loss accurately. To achieve fully quantized model, PyTorch internal operations (referred as aten ops)  have to be converted into quantized operations, and all dequantizers will be removed.

To implement PoT, we perform two steps, (1) Quantize the entire model according to above flow including weights and activations, while using a customized quantizer configuration function. (2) An algorithm loads all weights and convert them (restrict numbers) to be only power of two.

\section{Experiments / Results / Discussion}
\subsection{Performance Evaluation and Metrics}
We use cross-entropy loss and perplexity to measure the performance degradation due to quantization. To measure cross-entropy, we average over multiple iterations of calling the model (both floating and quantized) with input batches captured from testing set. Note that batches include inputs and corresponding targets. The cross-entropy $l_{ce}(t,y)$ with input $k-$dimension logits $t \in \mathbb{R}^k$ and target class $y$ is defined as follows,
\[ l_{ce}(t,y) = -\log P(y;t) = -\log \left( \frac{\exp(t_y)}{\sum_{s} \exp(t_s)} \right) \]

, and perplexity is just the exponential of the cross entropy defined as $\exp({l_{ce})} $

Additional metrics like ROUGE and BERTScore are highly recommended for future work to check generation quality. As will be mentioned in the results, cross-entropy might be misleading for unstable models and it is very useful to check generated text. In all trials, we quantize both linear layer weights and transformer tables, which results in huge memory and processing savings.

\subsection{Preliminary Character-level Quantized Performance}
Training is performed for 15000 iterations with 0.001 learning rate over Shakespeare training dataset, assuming all floating-point operations. Training loss is $\approx$ 1.45 while evaluation loss is $\approx$ 1.58. For testing, we tried quantize both wights/activations in in INT8 (8 bits, symmetric for weights and asymmetric for activations). The performance almost not changed in terms of cross entropy loss. Moreover, we tried limiting the integer range of weights to 5-bits (-16 to 15), and only very small loss is observed compared to floating point (1\% degradation in cross entropy).
Finally, we restricted weights to PoT in INT8 with symmetric range of $[-16,16]$ power levels (11 levels), and interestingly results are very good with cross entropy $\approx$ 1.89 which is very close to INT8 and 5-bits performance.

\subsection{GPT-2 Quantized Performance}
Extending Nano-GPT small model to much larger model GPT-2 was a bit challenging where multiple trials and tuning are performed to achieve properly working model.
Final results are shown in Table \ref{performance-table}\footnote{note that for normal quantization (2's complement implementation), the maximum supported positive $2^{N_{bits}-1}-1$, but we ignored "-1" in below integer ranges for simplicity}, where the base floating point cross entropy loss is $\approx 3.167$ (perplexity $\approx 23.733$). We tested fully quantized model with uniform quantization in INT8, INT16, and INT32 for weights (activations are kept in INT32). All formats show very good performance with little performance loss for INT8. However, limiting range to lower than INT8, like 5-bits is no longer an option, and is showing much higher degradation ($\approx 7.7$ cross entropy loss).

\begin{table}
\caption{Performance summary of Normal quantization versus PoT}
\label{performance-table}

\begin{tabular}{ |p{2cm}||p{1.5cm}|p{1.5cm}|p{1.5cm}| p{1.5cm}| p{1.5cm}| p{1.5cm}|}
 \hline
 & \multicolumn{3}{c|}{Normal quantization } & \multicolumn{3}{c|}{Power of Two (PoT)} \\
 \hline
 Integer range & No. of bits & $l_{ce}$  &Perplexity &No. of bits& $l_{ce}$& Perplexity\\
 \hline
 $[-2^{5},2^5]$   & 6 bits    &7.7&   2225.86  & 4 bits & 9.563 & 14158.84\\
 $[-2^{7},2^7]$&   8 bits  & 3.329   &27.92 & 4 bits & 4.5 & 90\\
 $[-2^{15},2^{15}]$ & 16 bits & 3.1888&  24.25 & 5 bits & 4.3 & 73.7\\
 $[-2^{20},2^{20}]$    & 32 bits & 3.1883&  24.24& 6 bits & 4.23 & 68.7\\
 $[-2^{30},2^{30}]$&   32 bits  & 3.187 & 24.23 & 6 bits & 4.08 & 59.14\\
 \hline
\end{tabular}
\end{table}

\begin{figure}[ht] 
    \caption{Output generated text with 4 bits PoT}
    \label{text}
    \centering
    \includegraphics[width=1\textwidth]{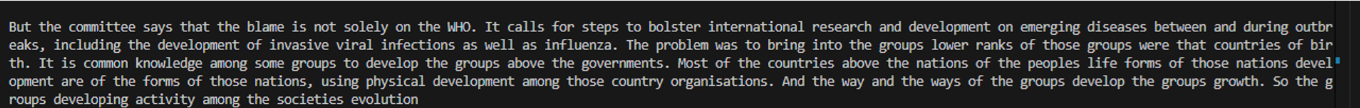}
\end{figure}

Considering PoT quantization, we observe cross entropy loss to degrade to $\approx$[4.08-4.5] based on the range limit. This slightly increase in loss is expected, in fact, it is very promising given that we did not perform any training, i.e., we only quantize weights after training (post training quantization PTQ not quantization assisted training QAT). In the following, we summarize very interesting and promising observations,
\begin{itemize}
    \item The PoT quantized model has very promising performance even at low number of power levels  (15 levels [-7,7]), which requires only 4-bits. This results in huge reduction of memory requirements (a factor of 32/4=8). 
    \item More interestingly, at same number of bit requirement (4-bits), normal uniform quantization shows a big failure (cross entropy $\approx$ 7.7) and completely off generated text, while PoT shows only $\approx 4.5$ cross entropy loss, with very good quality output text shown in figure \ref{text}. This can be expected given that PoT is non-uniform (log) quantization which can fit better the data with more levels for smaller values and less levels for outliers.
    \item An important factor that we discovered during tuning the PoT quantization parameters is epsilon $\epsilon$. This parameter is used to limit floating point operations resolution\footnote{remember that our quantized model only simulates quantization loss, but calculations will be done in floating point after dequantization operation} and ensure stable results (e.g., avoid divide by zero). Increasing $\epsilon$ too much may lead to more instability while decreasing too much will limit quantization resolution. Hence, $\epsilon$ should be properly selected based on range of quantized data. It is worth mentioning that choice of $\epsilon$ has more impact on the PoT compared to the normal quantization. Another note is that non-stable results might not be clear in cross entropy loss, but is very clear in generated text.
    \item Finally, the most important complexity reduction factor is replacing multiplication to bit shifting. This reduces all linear and transformer operations by factor of $\approx$4-5 since multiplication requires approximately 5 clock cycles while bit shifting is just one clock cycle.
\end{itemize}

\section{Conclusion / Future Work}
We proposed a very efficient way to reduce LLM complexity through power of two quantization for weights. The proposed technique is tested on GPT-2 124 M and is shown to achieve promising performance even with only 15 levels of powers (4-bits). This results in memory reduction with a factor of 8 and processing power reduction with a factor of 5. Our future work includes enhancing the quantized PoT weights model performance considering training QAT\footnote{note that in order to test the QAT, small LLM model is not suitable as it does not show much performance degradation, but we will need to consider GPT-2 training which requires large processing power}. Moreover, it will be very interesting to try bigger models like Llama with 1 B parameters as it recently gained more attention in edge devices. Finally, considering other performance metrics such as ROUGE and BERTScore.

\section{Appendix A}

\begin{figure}[H] 
    \caption{Snapshot of quantized graph of GPT-2 model}
    \label{graph}
    \centering
    \includegraphics[width=1.2\textwidth]{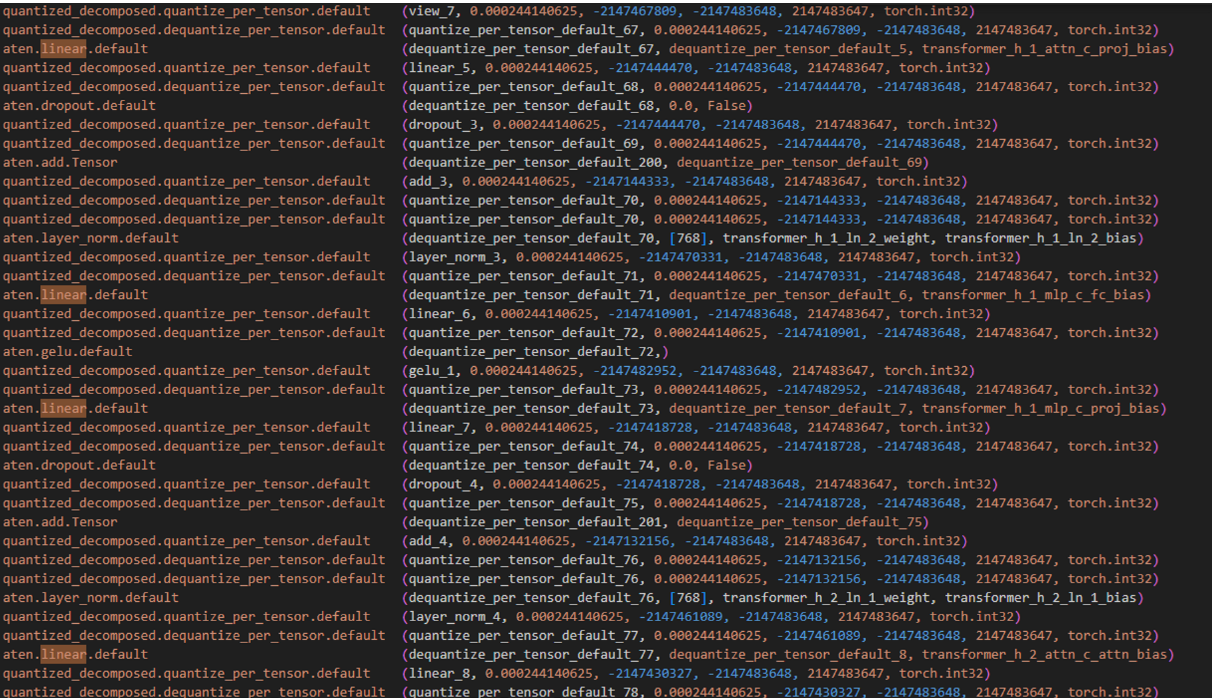}
\end{figure}

\begin{figure}[H] 
    \caption{Snapshot of Shakespeare dataset and encoded data}
    \label{Shak}
    \centering
    \includegraphics[width=1\textwidth]{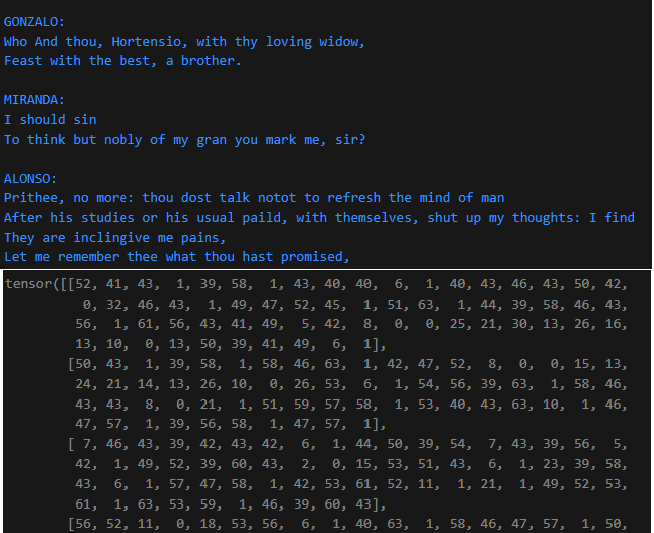}
\end{figure}

\begin{figure}[H] 
    \caption{Snapshot of OpenWebText dataset and encoded data}
    \label{Open}
    \centering
    \includegraphics[width=1\textwidth]{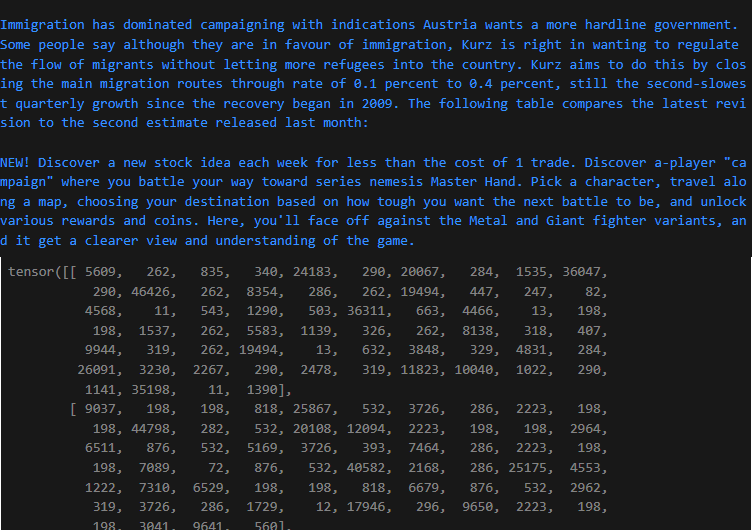}
\end{figure}

Figure \ref{graph} shows snapshot of quantized graph of GPT-2 model.
Figure \ref{Shak} shows a snapshot of Shakespeare dataset and encoded data.
Figure \ref{Open} shows a snapshot of OpenWebText dataset and encoded data.

\section{Contributions}
I have done all work myself, Mahmoud Elgenedy, and no other team members.
I would like to acknowledge my Instructors and TAs of CS229 for helping me significantly understand deep learning and LLMs.
I would like to acknowledge Marc Hoffman and Srijesh Sudarsanan for helping me understanding PyTorch quantization flow.
I would like to acknowledge useful discussion with TA Zikui Wang, and TA comments on projects submissions.
I got some help in basics questions from internet search and using LLMs as well as Google Colab.

\bibliographystyle{ieeetr}
\bibliography{references}

\end{document}